\documentclass[superscriptaddress,twocolumn,aps,prb,amsmath,floatfix]{revtex4}
\usepackage{graphicx}
\usepackage{dcolumn}
\usepackage{bm}

\newcommand{\ve}[1]{\mathbf{#1}}

\begin{document}

\title{Boundary Lubrication: Squeeze-out Dynamics of a Compressible 2D Liquid}
\author{U. Tartaglino}
\affiliation{INFM Democritos National Simulation Center, and Unit\`a INFM,
 Trieste}
\affiliation{SISSA, Via Beirut 4, I--34014 Trieste, Italy}
\author{B.N.J. Persson}
\affiliation{IFF, FZ-J\"ulich, 52425 J\"ulich, Germany}
\affiliation{ICTP, Strada Costiera 11, I--34014 Trieste, Italy}
\author{A.I. Volokitin}
\affiliation{IFF, FZ-J\"ulich, 52425 J\"ulich, Germany}
\affiliation{Samara State Technical University, 443100 Samara, Russia}
\author{E. Tosatti}
\affiliation{SISSA, Via Beirut 4, I--34014 Trieste, Italy}
\affiliation{INFM Democritos National Simulation Center, and Unit\`a INFM,
 Trieste}
\affiliation{ICTP, Strada Costiera 11, I--34014 Trieste, Italy}

\begin{abstract}
The expulsion dynamics of the last liquid monolayer 
of molecules confined between two surfaces has been 
analyzed by solving the two-dimensional (2D) Navier-Stokes
equation for a compressible liquid. We find that the 
squeeze-out is characterized by the parameter 
$g_0 \approx P_0/\rho c^2$, where $P_0$ is the average
perpendicular (squeezing) pressure, $\rho$ the liquid 
(3D) density and $c$ the longitudinal sound velocity 
in the monolayer film. When $g_0 \ll 1$ 
the result of the earlier incompressible treatment is 
recovered. Numerical results for the squeeze-out time, and for 
the time-dependence of the radius of the squeezed-out 
region, indicate that compressibility effects may be
non-negligible both in time and in space. In space, they 
dominate at the edge of the squeeze-out region. In time,
they are strongest right at the onset of the squeeze-out process, 
and just before its completion. 
 
\vskip 0.5cm \noindent 81.40.Pq, 46.55.+d, 68.35.Af, 62.20.Qp\vskip
0.5cm
\end{abstract}
\maketitle

\section{Introduction}
\label{sec:introduction}

Sliding friction is one of the oldest problems in physics, 
and has undoubtedly a huge practical 
importance\cite{P1a,P1b,P1c}. In recent years, the 
ability to produce durable low-friction surfaces and 
lubricants has become an important factor in the miniaturization 
of moving components in technologically advanced devices. 
For such applications, the interest is focused on the 
stability under pressure of thin lubricant films, 
since the complete squeeze-out of the lubricant from an 
interface may give rise to cold-welded junctions, 
resulting in high friction and catastrophically large wear.

It has been shown both experimentally and theoretically 
that when simple fluids  
(quasi-spherical molecules and linear hydrocarbons) 
are confined between atomically 
flat surfaces at microscopic separations, the behavior of 
the lubricant is mainly determined 
by its interaction with the solids that induce layering 
in the perpendicular direction 
\cite{Israel3a,Israel3b,Gao97a,Gao97b,Demi96a,Demi96b,Kleina,Kleinb}. 
The thinning of the lubrication film under applied pressure
occurs step-wise, by expulsion of individual layers. 
These layering transitions appear to be thermally 
activated \cite{PT,Persson2000,zilberman2001x1,zilberman2001x2}. 
Under strong confinement conditions, some lubricant fluids  become solid-like 
\cite{Israel3a,Israel3b,Gao97a,Gao97b,Demi96a,Demi96b,Kleina,Kleinb}. 
Other fluids, notably water \cite{KleinW,GranickW}, 
remain liquid-like up to the last layer that can be 
removed upon squeezing. 
This is related to the expansion of water upon freezing.\cite{Jagla,PerJCP},
and should also hold for other liquids which expand upon freezing.

The phenomenology of layering transitions in 2D-solid-like boundary  
lubrication has been studied in Ref.~\cite{PerJCP,Perssona,Perssonb}. 
It has been shown in a series of computer simulations that for 
solid-like layers, the layering transitions are sometimes 
initiated by a disordering transition, 
after which the lubricant behaves in a liquid-like manner for 
the rest of the squeeze-out process. 
Since the typical lateral extension in surface force apparatus (SFA) 
experiments is of order 10-100 $\mu$m (much greater than atomic dimensions), 
it is reasonable to expect that  {\em during the layering transition} 
the squeeze-out can be described in the framework of 2D continuum 
fluid mechanics.

For the  first time such squeeze-out layering transitions were 
quite recently observed in a chain alcohol, C$_{11}$H$_{23}$OH 
(Ref.~\cite{Mugele,Mugelesubm}), by imaging the 
lateral variation of the gap between the two anvil surfaces 
as a function of time.
These experiments addressed the $n=1 \rightarrow 0$ transition. 
More recently, in a refined experimental setup, Mugele et 
al.\ \cite{MugelToBe} were able to image several layering 
transitions ($n\rightarrow n-1$, $n=5,4,3$) of the silicon oil 
OMCTS (spherical molecule, diameter $\sim 10$ \AA) in great detail.

\begin{figure}[htbp]
    \leavevmode
    \scalebox{0.425}{\includegraphics{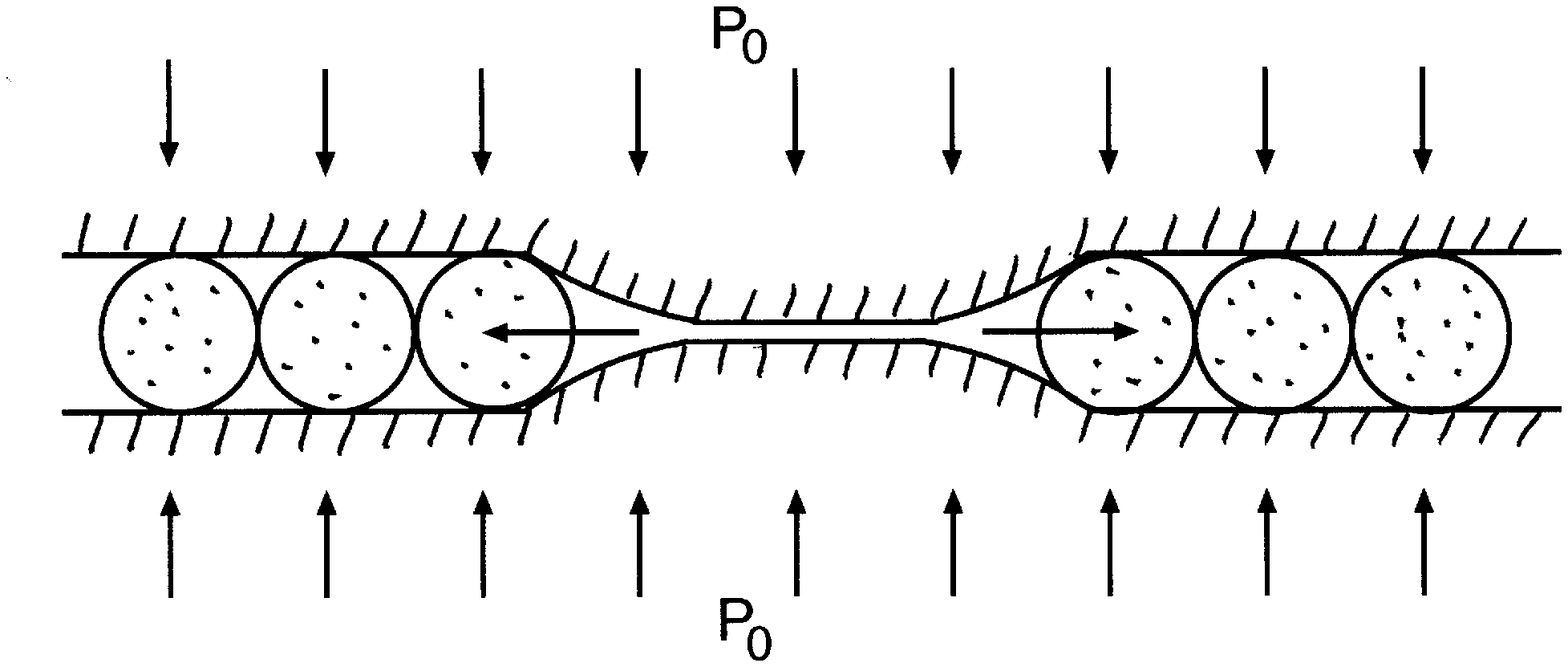}}
  \caption{\label{P0} Because of the curvature of the solid 
walls at the boundary line, the
perpendicular pressure $P_0$ will give rise to
a parallel force component acting on the 2D-lubrication film. (Schematic.)}
\end{figure}

The basic theory of 2D squeeze-out dynamics was outlined in Ref.~\cite{PT}.
Initially the system is trapped in a meta-stable state at the 
initial film thickness. Squeeze-out starts 
by a thermally activated nucleation process in which a density 
fluctuation forms a small hole, of critical radius $R_c\sim 10$ \AA. 
Once formed, a 2D pressure difference $\Delta p$ develops between the 
boundary line separating the squeezed out region from the rest of the system, 
and the outer (roughly circular) boundary line of the contact area,
thus driving out the rest of the 2D fluid. 
The origin of $\Delta p$ is the elastic relaxation of the 
confining solids at the
boundary line as is illustrated in Fig.~\ref{P0}.

All earlier analytical studies of squeeze out have 
assumed that the lubricant behaves as an {\it incompressible} 2D liquid. 
While that assumption is quite good for many practical situations,
recent computer simulations\cite{PerJCP} have shown that, at 
least at high squeezing pressures, strong density fluctuations 
may occur in the lubrication film. For example, in Ref.~\cite{PerJCP} 
it was found that during the layering transition $n=2\rightarrow 1$, while
islands of (temporarily) trapped bilayer ($n=2$) were removed 
by being squeezed into the monolayer,
the density of the monolayer film was much higher in the region
close to the trapped
$n=2$ islands than further away. This resulted in a
2D pressure gradient in the film which induced a flow of the 
lubricant molecules away from the trapped islands.
This kind of situations clearly calls for a consideration of the finite
compressibility of the film. 

In this paper we study the dynamics of the expulsion of the 
last liquid monolayer of molecules confined
between two surfaces by solving the two-dimensional (2D) Navier-Stokes
equation for a {\it compressible} liquid monolayer. We find 
that the squeeze-out is characterized by the parameter $g_0 
\approx P_0/\rho c^2$, where $P_0$ is the average
perpendicular (squeezing) pressure, $\rho$ the liquid (3D) density and $c$ the
longitudinal sound velocity in the monolayer film. When $g_0 \ll 1$ the
2D liquid can be considered as incompressible, in which case 
the results of the earlier treatment are reproduced. We present 
numerical results for the squeeze-out time, and for
the time-dependence of the radius of the squeezed-out region, 
for several values of the parameter $g_0$. 
The main changes due to compressibility occur right at the 
onset of the squeeze-out process, and just before its completion.

\section{Theory}
\label{sec:theory}

We assume the lubricant film to be in a 2D-liquid-like state, and the
squeeze-out to be described by the 2D Navier-Stokes equations. For a
compressible 2D liquid these equations take the form
$${\partial n \over \partial t}+\nabla \cdot (n \ve{v})=0\eqno(1)$$
$${\partial \ve{v}\over \partial t} + \ve{v}\cdot \nabla \ve{v}
 = 
-{1\over mn}\nabla p$$
$$ +\nu_1\nabla^2\ve{v}
+\nu_2\nabla \nabla\cdot \ve{v}-\bar \eta \ve{v}\eqno(2)$$
where $mn$ and $\ve{v}$ are the local 2D mass density and the velocity
of the fluid, $p$ is the 2D-pressure, $\nu_1$ and $\nu_2$ are viscosities. 
The last term in (2), i.~e.\ $-\bar{\eta}\ve{v}$, describes the drag
force acting on the fluid as it slides relatively to the solid
walls.\cite{slidingfriction} The magnitude of the friction also depends
on the nature of the solid walls, e.~g.\ on the amplitude of the
atomic corrugation and on the structure of the solid walls (amorphous
vs.\ crystalline, commensurate vs.\ incommensurate).
As long as the lubricant can be considered as a 2D fluid,
as in the recent squeeze out experiments by Mugele et al.,
all the details of the solid walls are properly taken into account
by the friction coefficient $\bar{\eta}$.
In principle $\bar{\eta}$ depends on the normal pressure too,
but in the measurements of Mugele et al.\ this dependence is
slight enough that it appears to be negligible\cite{zilbermannew}.

Simple dimensional arguments (see Ref.~\cite{zilberman2001x2}) show 
that one can usually neglect the nonlinear and the viscosity 
terms in (2), and that one can also assume the velocity field 
to change so slowly that the time derivative term can be 
neglected too. Thus,
 $$\nabla p +mn\bar \eta \ve{v}=0.\eqno(3)$$
In what follows we will assume that the squeeze-out nucleated 
in the center ($r=0$) of the contact area,
and spread circularly towards the periphery.
The 2D-pressure $p$ at the outer boundary $r=R$ of the contact 
area takes the constant value $p_0$  (the spreading pressure) 
while it takes a higher value $p_1$ at the inner
boundary towards the $n=0$ area (Fig.~\ref{P0}). In fact, if
$P(r)$ is the perpendicular pressure acting in the contact area, 
then $p_1(r) = p_0 +P(r)a$,
where $a$ is the thickness of the monolayer (see Ref.~\cite{PT}).
We may in most practical applications assume a Hertz contact pressure
$$P(r) = {3\over 2} P_0 \left (1-{r^2\over R^2}\right )^{1/2}\eqno(4)$$
Finally, in order to have a complete set of equations we must specify
the relation between the 2D pressure $p$ and the 2D density $n$. 
We will assume that $$p=p_0+mc^2(n-n_0)\eqno(5)$$
where the compressibility is $B=1/mc^2$, $c$ being the 
longitudinal 2D sound velocity. Here $n_0$ is the 2D lubricant 
density at the periphery $r=R$ of the contact area.

Writing the velocity field as $\ve{v} = \hat{r}v(t)$,
equations (1) and (3) takes the form
$${\partial n \over \partial t}+\left ({\partial \over \partial r}+{1\over r}\right ) (nv) =0\eqno(6)$$
$${\partial p \over \partial r}+mn\bar \eta v = 0$$
Using (5) the last equation takes the form
$${\partial n \over \partial r}+{\bar \eta \over c^2} (nv) = 0\eqno(7)$$
Combining (6) and (7) gives
$${\partial n \over \partial t} - {c^2\over \bar \eta} 
\left ({\partial \over \partial r} +{1\over r}\right )
{\partial n \over \partial r}=0\eqno(8)$$
The density $n$ satisfies the boundary conditions
$$n(R,t)=n_0\eqno(9)$$
$$n(r_1(t),t)=n_0+{a\over mc^2} P(r_1(t))\eqno(10)$$
where $r=r_1(t)$ is the equation for the inner boundary line. 
Finally, we note that the velocity $v(r_1(t),t)$ of the 
2D-liquid at the inner boundary $r=r_1(t)$ must equal
the radial velocity $\dot r_1(t)$ of the boundary line. 
Thus, if we put $r=r_1(t)$ 
and $v(r_1(t),t)=\dot r_1(t)$ in (7) we get
$${\partial n\over \partial r}(r_1(t),t) = -{\bar \eta 
\over c^2} n(r_1(t),t)\dot r_1\eqno(11)$$
Let us at this point introduce dimensionless variables. 
If we measure the radius $r$ in units of $R$, time $t$ in units of
$\tau=\bar \eta R^2/c^2$ and density $n$ in units of $n_0$ we get

$${\partial n \over \partial t} - \left ({\partial \over \partial r} 
+{1\over r}\right ){\partial n \over \partial r}=0\eqno(12)$$

and the boundary conditions becomes
$$n(1,t)=1\eqno(13)$$
$$n(r_1(t),t)=1+ g(t)\eqno(14)$$
and
$${\partial n\over \partial r}(r_1(t),t) = - [1+g(t)]\dot r_1\eqno(15)$$
where 
$$g(t)={a\over mc^2 n_0} P(r_1(t))\eqno(16)$$
If we assume that $P(r)$ is of the Hertz form, then
$$g(t)=g_0{3\over 2}\left (1-[r_1(t)]^2\right )^{1/2}\eqno(17)$$
where
$$g_0 = {aP_0\over mc^2 n_0}.\eqno(18)$$
Thus, the theory depends only on a single parameter $g_0$.
Note that $n_0/a \approx \rho$, where $\rho$ is the 3D density 
of the liquid, and with the typical values $\rho \approx 1000 
\ {\rm kg/m^3}$ and $c \approx 700 \ {\rm m/s}$ we get $\rho c^2 \approx 500
\ {\rm MPa}$. In the experimental studies by Mugele and Salmeron (as well
as in most other Surface Force Apparatus studies) the average 
squeezing pressure $P_0 \ll 500 \ {\rm MPa}$ which implies 
that the liquid can be considered as
incompressible and the theory developed elsewhere can be 
used\cite{PT,zilberman2001x1,zilberman2001x2}. However, in many
practical situations the pressure $P_0$ might be similar to the yield stress
of the solids which for metals is typically of order $\sim 1000 \ {\rm MPa}$.
In these cases it is necessary to include the finite
compressibility of the lubricant in order to accurately 
describe the squeeze-out dynamics.
In Sec.\ III we shall present numerical results based on 
the above equations for a range of values of the parameter $g_0$. 

Let us first calculate the squeeze-out time in the limit 
when the compressibility $B=\infty$. In this limit, all the 
adsorbates from the region $r<r_1(t)$ will be piled-up right 
at the boundary line $r=r_1$. If we consider a
small angular section $\Delta \phi$ then the driving force acting on
the boundary line is $F=r_1 \Delta \phi [p_1-p_0] = r_1 \Delta \phi P(r_1) a$.
This must balance the frictional drag force which is 
$Nm\bar \eta \dot r_1$ where
the number of adsorbates $N=\pi r_1^2 (\Delta \phi/2\pi)n_0$. Thus we get
$$\dot r_1 r_1 = {2P(r) a \over n_0 m \bar \eta} \eqno(19)$$
Assume first that $P=P_0$ is constant. Thus, (19) gives
$$\dot r_1 r_1 = {R^2 \over 2 T}\eqno(20)$$
where
$$T={mn_0\bar \eta R^2 \over 4 P_0a}\eqno(21)$$
is the squeeze out time for an incompressible 2D fluid, see Ref.~\cite{PT}.
Integrating (20) gives
$${r_1(t)\over R} = \left ({t\over T}\right )^{1/2}\eqno(22)$$
so that the squeeze-out time, $T^*$, in the limit $B=\infty$ 
is the same as for an incompressible 2D-fluid, $T^*=T$.
This suggests that the squeeze-out time is {\it independent} 
of the compressibility of the 2D-liquid, which our numerical 
simulations presented below indeed show to be the case. 

Next, let us assume that the perpendicular pressure is of 
the Hertz form, appropriate for a fluid between curved surfaces. 
In that case (19) gives
$$\dot r_1 r_1 = {R^2\over 2 T}{3\over 2}\left [1-\left ( {r_1\over R}\right )^2\right ]^{1/2}$$
or, with $r_1^2/R^2=\xi$, 
$$\int_0^{r_1^2/R^2}{d\xi \over (1-\xi)^{1/2}} = {3t\over 2T}$$
Performing the integral gives
$$\left [1-\left ({r_1\over R}\right )^2 \right ]^{1/2} = 1- {3t\over 4T}$$
Thus, the squeeze-out time for the Hertzian contact pressure is
$T^*_{\rm H} = 4T/3$, and the time dependence of $r_1(t)$ is given by 
$${r_1\over R}=\left ({3t\over 2T}\right )^{1/2} \left (1-{3t\over 8T}\right )^{1/2}$$
or
$${r_1\over R}=\left ({2t\over T^*_{\rm H}}\right )^{1/2} \left (1-{t\over 2T^*_{\rm H}}\right )^{1/2}$$
The squeeze-out time for an incompressible fluid with a 
Hertzian contact pressure is
$T_{\rm H} = (4T/3)(2-{\rm ln}4)\approx 0.8183 \ T$ which is a factor $2-{\rm ln}4 
\approx 0.6137$ smaller than for the $B=\infty$ limiting case.
Thus, for a Hertzian contact pressure the squeeze-out time 
does depend on the compressibility of the 2D lubrication film.

\section{Numerical results}
\label{sec:numerical}

\begin{figure}[htbp]
    \leavevmode
    \scalebox{0.60}{\includegraphics{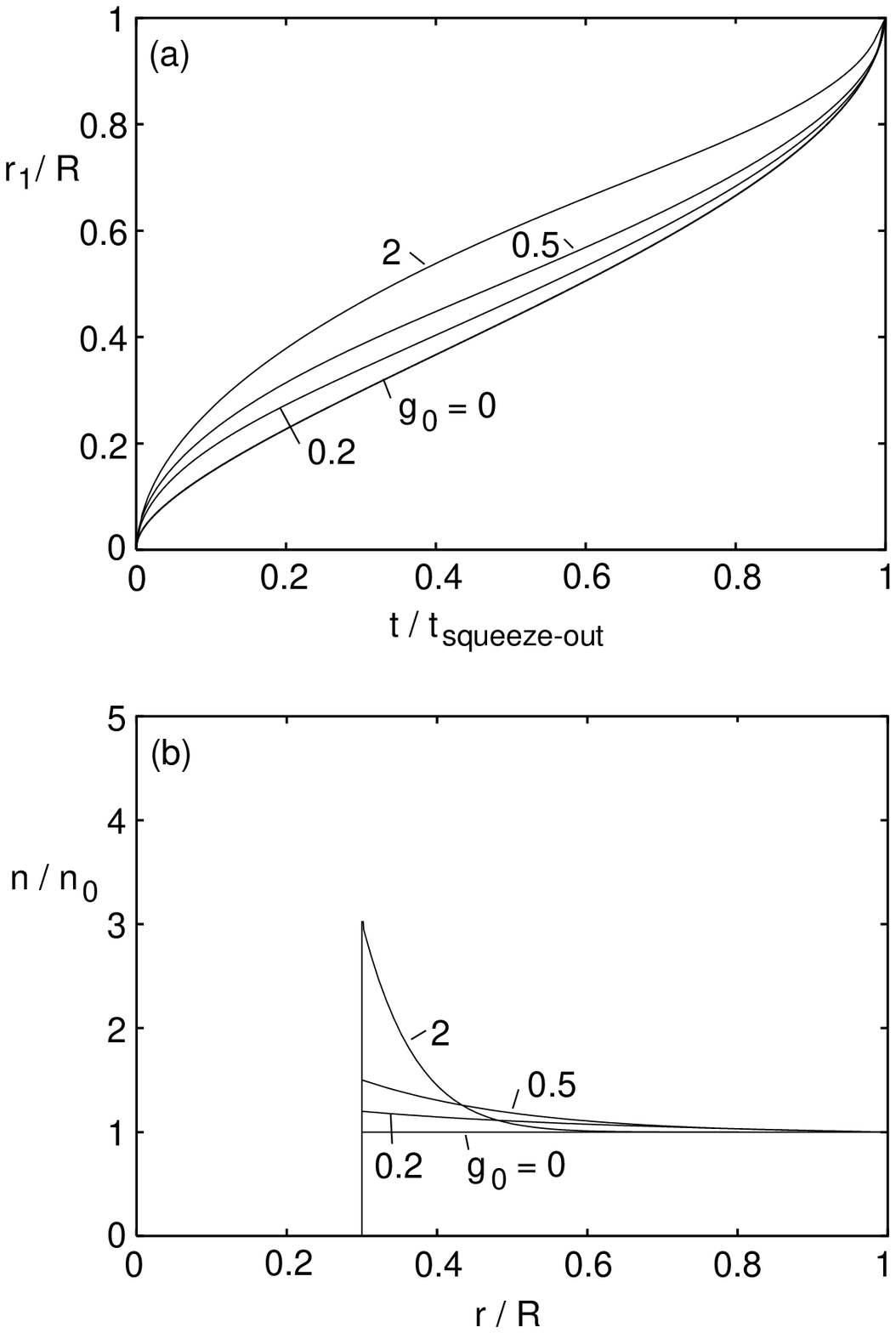}}
  \caption{\label{r1.t} Constant squeezing pressure, $P(r)=P_0$.
(a) Squeeze-out radius $r_1$ (in units of the radius $R$ of the contact area) 
versus time (in units of the squeeze out time). We 
show results for four different cases, namely $g_0=0$ (corresponding to an
incompressible adsorbate layer), $0.2$, $0.5$ and $2.0$.
(b) Adsorbate density distribution $n(r)$ (in units of the 
natural density $n_0$) during squeeze-out, for the same four cases 
as in (a) at a time when the squeeze out radius $r_1 = 0.3 R$. } 
\end{figure}

We consider first the case of a spatially constant squeezing pressure, 
$P(r)=P_0$, appropriate for a fluid between flat surfaces. 
In this case numerical calculations show that 
the squeeze-out time is independent of the compressibility,
i.e., independent of the parameter $g_0$. However, the 
time dependence of the squeeze-out radius
$r_1(t)$ does depend on $g_0$. In Fig. \ref{r1.t}(a) we show 
this time dependence for four cases, $g_0=0$, $0.2$, $0.5$ and $2$. 
In Fig. \ref{r1.t}(b) we show the adsorbate density
profile at the time point when the squeeze-out radius $r_1 = 0.3 R$, 
for the same four $g_0$-values as in (a).

\begin{figure}[htbp]
    \leavevmode
    \scalebox{0.60}{\includegraphics{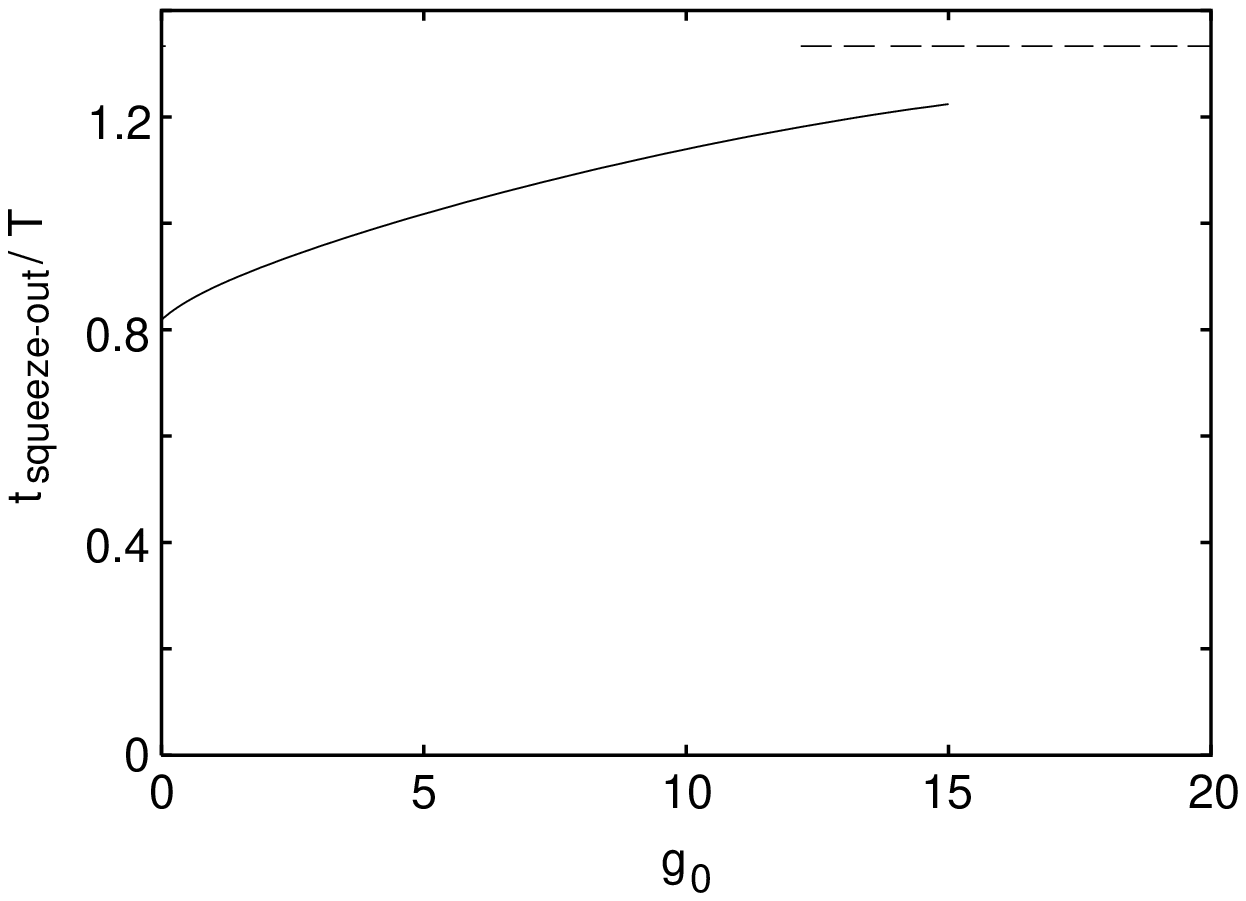}}
  \caption{\label{time.g0} Dependence of the squeeze-out time on the compressibility parameter
$g_0$ for a Hertzian squeezing pressure.} 
\end{figure}

\begin{figure}[htbp]
    \leavevmode
    \scalebox{0.600}{\includegraphics{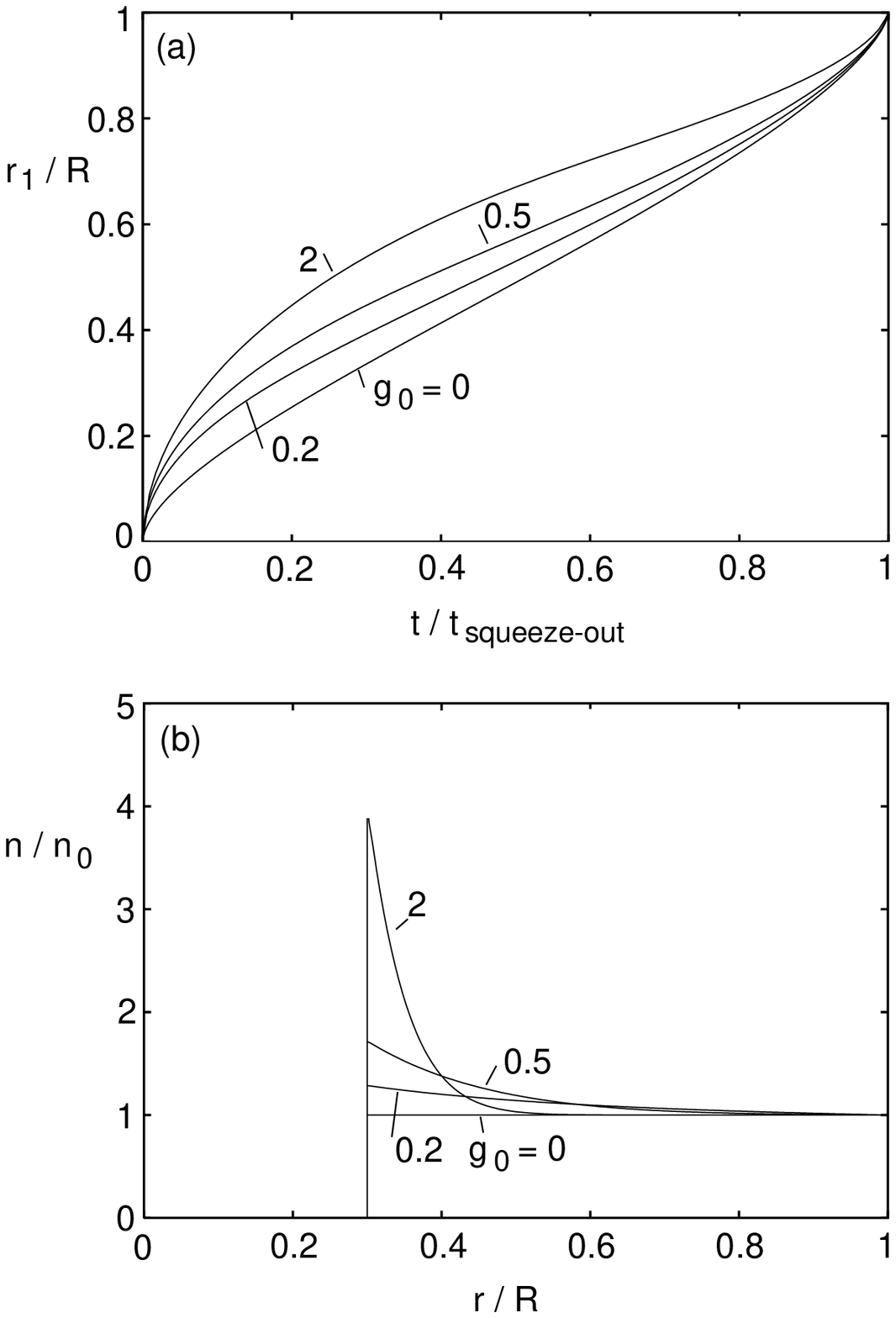}}
  \caption{\label{r1.t.Hertz} Hertzian squeezing pressure, 
$P(r)$ given by Eq.~(4).
(a) Squeeze-out radius $r_1$ (in units of the radius $R$ of the contact
area)
versus time (in units of the squeeze out time). We
show results for four different cases, namely $g_0=0$ (corresponding to an
incompressible adsorbate layer), $0.2$, $0.5$ and $2.0$.
(b) Adsorbate density distribution $n(r)$ (in units of the
natural density $n_0$) during squeeze-out, for the same four cases
as in (a) at a time when the squeeze out radius $r_1 = 0.3 R$. }
\end{figure}

Let us now assume the Hertzian squeezing
pressure profile, $P(r)=P_{\rm H}(r)$. In this case the squeeze-out time depends on the compressibility,
increasing from $\approx 0.8183 \ T$ to $1.3333 \ T$ as the compressibility increases from $B=0$ to $\infty$.
In Fig. \ref{time.g0} we show the squeeze-out time as a function of $g_0$. In Fig. \ref{r1.t.Hertz} we show
the  
time dependence of $r_1(t)$ for 
four cases, $g_0=0$, $0.2$, $0.5$ and $2.0$. In Fig. \ref{r1.t.Hertz}(b) we show the adsorbate density
profile at the time point when the squeeze-out radius $r_1 = 0.3 R$, for the same four
$g_0$-values as in (a). 
The results of Fig.~\ref{r1.t} and Fig.~\ref{r1.t.Hertz} show that a
compressibility parameter $g_0$ as small as 0.2 should produce a
measurable difference in the squeeze-out evolution at all times.
The main effect of compressibility appears at the beginning and at the
end of the squeeze-out process. Initially, compressibility
favors piling up of fluid at the squeeze-out boundary, which
can as a result expand more rapidly, compared with the
case of an incompressible fluid. On the other hand, when the hole approaches
the boundary of the contact region, the squeezing-out speed of the
compressed fluid is smaller, due to its increased density and, 
consequently, friction. In the case of uniform squeezing pressure 
these two effects compensate exactly, leading to a total squeeze-out 
time independent of the compressibility. For a Hertzian pressure 
distribution, the squeeze-out time instead increases with increasing
compressibility: the initial speed up is overcompensated by the
enhanced friction at the periphery of the contact area.

\section{Summary}
\label{sec:summery}

The continuum mechanics theory of 
squeeze-out has been solved numerically for 
a 2D {\it compressible} liquid.
We considered both a constant normal stress, and a 
Hertzian normal stress, and assumed a centro-symmetric squeeze-out. 
We found that the squeeze-out is completely 
characterized by the compressibility parameter 
$g_0 \approx P_0/\rho c^2$, where $P_0$ is the average
perpendicular squeezing pressure, $\rho$ the (3D) liquid density and $c$ the
longitudinal sound velocity in the monolayer film. When $g_0 \ll 1$ the
2D liquid can be considered incompressible, and the earlier results
are reproduced. We presented numerical results for 
the squeeze-out time, and for the time dependence of the 
squeeze-out radius, for a grid of values
of the parameter $g_0$. For a constant squeezing pressure, 
the squeeze-out time was found to be {\it independent} of the 
compressibility parameter $g_0$, while for a Hertzian contact pressure
it increased slightly with increasing $g_0$. It is hoped that
these theoretical results will soon be submitted to experimental check.

\begin{acknowledgments}
B.P. thanks BMBF for a grant related to the German-Israeli Project Cooperation 
``Novel Tribological Strategies from the Nano-to Meso-Scales'',
and ICTP/SISSA in Trieste where the main part of this work was carried out.
Work at SISSA was sponsored by INFM PRA NANORUB, by MIUR through COFIN 2001,
and by EU, contract ERBFMRXCT970155 (FULPROP).
\end{acknowledgments}

\end{document}